 \documentclass[preprint,twocolumn,3p]{elsarticle}
\usepackage{hyperref}
\usepackage{bm}
\usepackage{amsmath}
\usepackage{graphicx}%
\usepackage{amsfonts}%
\usepackage{amssymb}
\usepackage{color}
\usepackage{fullpage}
\usepackage{widetext}
\usepackage{bigints}
\usepackage{xfrac}

\biboptions{numbers,sort&compress}

\usepackage{afterpage}

\renewcommand*{\today}{April 10, 2022}

\begin{document}

\begin{frontmatter}

\title{The Dukhin number as a scaling parameter for selectivity in the infinitely long nanopore limit: extension to multivalent electrolytes} 

\author[1]{Zs\'ofia Sarkadi}
\author[2]{D\'avid Fertig}
\author[1]{M\'onika Valisk\'o}
\author[1]{Dezs\H{o} Boda\corref{cor}}
\cortext[cor]{Author to whom correspondence should be addressed. E-mail address: boda@almos.vein.hu}
\address[1]{Center for Natural Sciences, University of Pannonia, Egyetem u.\ 10, Veszpr\'{e}m, 8200, Hungary}
\address[2]{Laboratory of Engineering Thermodynamics, Technische Universit\"{a}t Kaiserslautern, 67663 Kaiserslautern, Germany}
\date{\today}


\begin{abstract}
Scaling of the behavior of a nanodevice means that the device function (selectivity, in this work) is a unique function of a scaling parameter that is an appropriate combination of the device parameters. Although nanopores facilitate the transport of ions through a membrane of finite length if the pore is long compared to the pore radius, we deal with an important limiting case, the infinitely long nanopore (nanotube). In this case, device parameters are the pore radius, the electrolyte concentration, the surface charge density on the nanopore’s wall, and ionic valences. While in our previous study (Sarkadi et al., J. Chem. Phys. 154 (2021) 154704.) we showed that the Dukhin number is an appropriate scaling parameter in the nanotube limit for 1:1 electrolytes, in this work we obtain the Dukhin number from first principles on the basis of the Poisson-Boltzmann (PB) theory and generalize it to electrolytes containing multivalent ions as well. We show that grand canonical Monte Carlo simulations for charged hard spheres in an implicit solvent give results that are similar to those obtained from the PB theory with deviations that are the consequences of ionic correlations (including finite size of ions) beyond the mean-field level of the PB theory. Such a deviation occurs when charge inversion is present, in 2:2 and 3:1 electrolytes, for example.  
\end{abstract}

\end{frontmatter}

\section{Introduction}
\label{sec:intro}

When the function of a device is determined by a few well-defined input parameters, $a_{1}, a_{2}, \dots$, it is often possible to group them into a composite parameter, $\xi$, that determines the device's behavior by itself.
This scaling parameter is a simple analytical function of the independent variables: $\xi=\xi(a_{1}, a_{2}, \dots)$.
Let $F$ be the device function, an observable property of the device obtained as an output of the measurement or the calculation.
Scaling of the device function means that $F$ is a smooth unambiguous function of the scaling parameter: $F=f\left[ \xi(a_{1}, a_{2}, \dots) \right]$.

This behavior can have a practical importance because we can predict the device function, $F$, for a combination of the input parameters, $\{a_{1}, a_{2}, \dots\}$, if we know the device function, $F'$, for another parameter set, $\{a'_{1}, a'_{2}, \dots\}$, that gives the same scaling parameter, $\xi(a_{1}, a_{2}, \dots)=\xi(a'_{1}, a'_{2}, \dots)$.
In this case, the device functions are also the same, $F=F'$, provided that the $f\left[ \xi \right]$ relationship exists.
This way, we can save possibly expensive measurements for a portion of the parameter space.

If we consider, for example, a nanopore in a membrane connecting two baths and facilitating the controlled transport of ions between them, we can identify the input device parameters as the radius and length of the pore, $R$ and $H$, the voltage imposed across the membrane, $U$, the concentration of the electrolyte, $c$, the surface charge pattern on the wall of the nanopore, $\sigma$, and ionic valences, $z_{+}$ and $z_{-}$.

The device function is often a dimensionless number expressing the relation between various output quantities.
An important example is the number that characterizes the ratio of the surface and volume conductances.
The idea of this parameter goes back to Bikerman \cite{bikerman_1940}.
A series of authors~\cite{wiersema_jcis_1966,obrien_jcsf_1978,obrien_cjc_1981,dukhin_advcollsci_1993,lyklema_csa_1998} adopted the idea to study electrophoretic phenomena.
Dukhin himself called this number $\mathrm{Rel}$.~\cite{dukhin_advcollsci_1993}
It was Lyklema who introduced the name `Dukhin number' to salute Dukhin.~\cite{lyklema_book_1995}
Its original definition is
\begin{equation}
 \mathrm{Du}^{\mathrm{out}} = \frac{\kappa^{\sigma}}{\kappa^{\mathrm{b}}h},
 \label{eq:duout}
\end{equation} 
where $\kappa^{\sigma}$ is the surface conductance, $\kappa^{\mathrm{b}}$ is the volume (bulk) conductance, and $h$ is a distance parameter characteristic of the geometry at hand, the radius of the colloid particle or the pore, for example.
The superscript `out' indicates that the conductances, $\kappa^{\sigma}$ and $\kappa^{\mathrm{b}}$, are output results of the measurements or calculations. 
As a consequence, $\mathrm{Du}^{\mathrm{out}}$ can be considered a device function.

In accordance with our concept of scaling the need to develop a simple parameter depending on the input device parameters and, in the meantime, characterizing the device's behavior also arose. 
A parameter called Bikerman-Dukhin number, $\mathrm{Bi}$,~\cite{bazant_pre_2004,chu_pre_2006} or just Dukhin number, $\mathrm{Du}$,~\cite{khair_jfm_2008,das_langmuir_2010,bocquet_chemsocrev_2010,zangle_csr_2010,lee_nanolett_2012,yeh_ijc_2014,ma_acssens_2017,xiong_scc_2019,poggioli_jpcb_2019,dalcengio_jcp_2019,kavokine_annualrev_2020,noh_acsnano_2020} was proposed as the ratio of the excess counterion quantity in the double layer and the quantity of ions in the bulk.
For a 1:1 electrolyte, if we assume perfect exclusion of coions, the excess counterion quantity is proportional to $ |\sigma| 2\pi R H$, while the quantity of charge carriers in the  bulk electrolyte is proportional to $2c R^{2}\pi H$. 
Their ratio is   
\begin{equation}
 \mathrm{Du}^{1:1} \equiv \frac{|\sigma|}{eRc} .
 \label{eq:dukhin}
\end{equation} 
This expression contains input device parameters, $\sigma$, $R$, and $c$, as opposed to $\mathrm{Du}^{\mathrm{out}}$ in Eq.~\ref{eq:duout}  ($e$ is the elementary charge).
This number is a limiting value because perfect anion exclusion is an approximation, still, it may characterize the device behavior at other conditions as well.
If scaling prevails, $\mathrm{Du}^{1:1}$ may serve as a scaling parameter that in itself may determine some device function through a well-defined $f[\mathrm{Du}^{1:1}]$ relationship.

Indeed, in our previous work~\cite{sarkadi_jcp_2021}, we showed that for the infinitely long ($H\to \infty$) uniformly charged nanopore, cation selectivity defined as
\begin{equation}
 S_{+}^{I}=\frac{I_{+}}{I_{+}+I_-} 
 \label{eq:sel}
\end{equation} 
scales with $\mathrm{Du}^{1:1}$. 
In this equation,  $I_{i}$ is the electrical current that is related to the particle current through $I_{i}=z_{i}eJ_{i}$.
Depending on the sign of $\sigma$, $S^{I}_{+}$ is $1$ for perfect cation selectivity, $0.5$ for the non-selective case ($I_{+}=I_{-}$), and $0$ for perfect anion selectivity (here, we assume that the mobilities of the ions are the same).

Just as $\mathrm{Du}^{\mathrm{out}}$, selectivity is also composed of quantities that are outputs of calculations or measurements ($I_{+}$ and $I_{-}$).
What is more, $\mathrm{Du}^{\mathrm{out}}$ and $S^{I}_{+}$ are interrelated.
If surface conductance dominates, the pore is selective for the counterion.
If bulk conductance dominates, cations and anions contribute to the current equally, so the pore is non-selective.
Selectivity, however, is better suited for modeling studies, because it can be related to concentration profiles that are the main results of the calculations.

Other device functions may also be used for other charge patterns such as rectification for bipolar nanopores~\cite{hato_pccp_2017,fertig_jpcc_2019,fertig_pccp_2020,fertig_ms_2022} and switching for transistor-like patterns~\cite{madai_pccp_2018,fertig_mp_2018}.
In our previous works,~\cite{madai_pccp_2018,fertig_jpcc_2019,sarkadi_jcp_2021,fertig_ms_2022} we reported scaling behaviors for these nanopores. 
While most of these nanopores were finite in length, in our previous study~\cite{sarkadi_jcp_2021} we changed the $H/R$ aspect ratio and obtained the nanohole geometry ($H/R\to 0$) and the nanotube geometry ($H/R\to \infty$) as limiting cases.

In this study, we take a closer look at the nanotube limit and show that the Dukhin number defined in Eq.~\ref{eq:dukhin} can be obtained as a limiting value to the selectivity on the basis of the Poisson-Boltzmann (PB) theory.
We prove that it is indeed a scaling parameter for selectivity, and generalize it for multivalent electrolytes.

We find that scaling is more accurate using a simplified mean-field theory (PB); Monte Carlo (MC) simulations that take ionic correlations beyond the mean-field approximation of the PB theory into account produce deviations from the mean-field scaling.
In this respect, these deviations can be considered as indicators of the importance of strong electrostatic or volume exclusion correlations.

\section{Theory}

Because the basic function of a nanopore/membrane system is that it facilitates controlled transport of ions through the membrane, it is evident that we express composit parameters in terms of transport properties as in Eq.~\ref{eq:duout} and selectivity in terms of currents as in Eq.~\ref{eq:sel} (also called permselectivity).

In this work, however, we deal with the behaviors of ionic profiles at varying conditions using equilibrium computational methods.
Therefore, we intend to express the selectivity in terms of the equilibrium amounts of the competing ionic species in a unit length of the pore computed as 
\begin{equation}
 N_{i}= \int_{0}^{R} c_{i}(r) 2\pi r \mathrm{d}r , 
\label{eq:N_i}
 \end{equation} 
where $c_{i}(r)$ is the radial concentration profile.
This quantity may also be termed as occupancy or adsorption. 

\subsection{Definitions of selectivity}

We relate the two kinds of selectivity to each other on the basis of the Nernst-Planck (NP) equation. 
Transport flows only in the $z$ direction, so the NP equation is expressed for the $z$-component of the flux density as
\begin{equation}
 j_{i,z}(r)= -\frac{D_{i}}{kT} c_{i}(r) \left( \frac{\partial \mu_{i}(z,r)} {\partial z} \right)_{r} = \frac{D_{i}}{kT} c_{i}(r) z_{i}e E_{z},
 \label{eq:np}
\end{equation} 
where  $k$ is Boltzmann's constant, $T$ is the absolute temperature (it is $298.15$ K in this work), $D_{i}$ is the diffusion constant, and $\mu_{i}(z,r)$ is the (electro)chemical potential profile for ionic species $i$.
We assume that $(\partial \mu_{i}/\partial z )_{r}= -z_{i}eE_{z}$ is constant, where $E_{z}$ is the applied electric field.
We also assume that $D_{i}$ is constant.

Using Eqs.~\ref{eq:N_i} and \ref{eq:np}, the total particle current of species $i$ is obtained as
\begin{equation}
 J_{i}=\int_{0}^{R} j_{i,z}(r)2\pi r \mathrm{d}r = \frac{eE_{z}D_{i}}{kT} z_{i} N_{i} 
 \label{eq:J-vs-N}
\end{equation} 
so $|J_{i}|$ is proportional to $D_{i}|z_{i}|N_{i}$. 

We have multiple choices for constructing a composite parameter that characterizes selectivity.
Rational functions as in Eq.~\ref{eq:sel} are advantageous because they are bounded.
Instead of Eq.~\ref{eq:sel}, we use 
\begin{equation}
 S_{+}=\dfrac{|J_{+}|-|J_{-}|}{|J_{+}|+|J_{-}|}
\end{equation} 
because its value is zero in the non-selective case even for multivalent electrolytes provided that $D_{+}=D_{-}$.
We relate this $J_{i}$-based definition of $S_{+}$ to the $N_{i}$-based definition via Eq.~\ref{eq:J-vs-N} and obtain that
\begin{equation}
 S_{+}=\frac{D_{+}z_{+}N_{+}-D_{-}|z_{-}|N_{-}}{D_{+}z_{+}N_{+}+D_{-}|z_{-}|N_{-}} .
 \label{eq:sel2}
\end{equation} 
Next, we discuss how to compute $N_{i}$.

\subsection{Model and methods}

Let us consider the limit of the infinitely long cylindrical nanopore ($H\to  \infty$) of radius $R$ carrying uniform $\sigma$ surface charge density on its wall.
It is an idealization of a nanopore whose length is much larger than its radius ($H/R\gg 1$).
Concentrations are the same on the two sides of the membrane.
For the $H\to \infty$ limit it means that the system inside the pore is in equilibrium with a bath of fixed concentration (fixed chemical potential).

For the stoichiometric coefficients, it holds that $\nu_{+}z_{+}+\nu_{-}z_{-}=0$, while they are relative primes and positive integers ($\nu_{+}=\nu_{-}=1$ for symmetric electrolytes).
The ionic bulk concentrations are then related to the salt concentration through $c^{\mathrm{b}}_{i}=\nu_{i}c$.
These concentrations should be considered in the unit of number of particles in unit volume, but later in the Results section we will report them in the unit of mol/dm$^{3}$.

In our previous work~\cite{sarkadi_jcp_2021}, we rewrote the Dukhin number in the form
\begin{equation}
 \mathrm{Du}^{0}= - \frac{\sigma 8\pi l_{\mathrm{B}}\lambda_{\mathrm{D}}^{2}}{eR} ,
 \label{eq:du1}
\end{equation} 
where 
\begin{equation}
\lambda_{\mathrm{D}} = 
\left( \dfrac{ c e^{2}}{\epsilon_{0}\epsilon kT} \sum_{i} z_{i}^{2}\nu_{i} \right)^{-1/2} ,
\label{eq:lambdaD}
\end{equation} 
is the Debye length, $\epsilon$ is the dielectric constant of the solvent (it is $78.45$ in this work), $\epsilon_0$ is the permittivity of vacuum, and $l_{\mathrm{B}}=e^{2}/4\pi \epsilon_0 \epsilon kT$ is the Bjerrum length.
The negative sign is relevant for asymmetric electrolytes, where the device has different behaviors for different signs of $\sigma$.
This choice allows that $S_{+}>0$ for $\mathrm{Du}^{0}>0$.
Note that $\mathrm{Du}^{0}=\mathrm{Du}^{1:1}$ only for 1:1 systems.

We use two statistical mechanical methods for this model of the nanopore.~\cite{sarkadi_jcp_2021}
One is solving the PB equation 
\begin{equation}
 \frac{1}{r} \frac{\mathrm{d}}{\mathrm{d} r} \left( r \frac{\mathrm{d} \Phi(r)}{\mathrm{d} r} \right) = 
 -\frac{e}{\epsilon_0 \epsilon} \sum_{i=1}^{2} z_{i}c^{\mathrm{b}}_{i} e^{-z_{i}e\Phi(r)/kT}
 \label{eq:pb}
\end{equation} 
for the mean electrical potential profile, $\Phi(r)$, with the boundary conditions that $\left(\mathrm{d}\Phi(r)/\mathrm{d}r\right)_{r=0}=0$ and $\Phi(0)=\Phi_{0}$. 
The surface charge follows from
\begin{equation}
\left( \frac{\mathrm{d}\Phi(r)}{\mathrm{d}r} \right)_{r=R} = \frac{1}{\epsilon_{0}\epsilon} \sigma , 
\label{eq:sigma}
\end{equation} 
while the concentration profiles are obtained as 
\begin{equation}
c_{i}(r)=c^{\mathrm{b}}_{i}\exp\left( -z_{i}e\Phi(r)/kT \right) .
\label{eq:c_i}
\end{equation} 

We also solve the problem with the grand canonical MC simulation method, where insertions/deletions of neutral ion clusters ($\nu_{+}$ cations and $\nu_{-}$ anions) are performed for a fixed chemical potential of the salt, $\mu_{\pm}=(\nu_{+}\mu_{+}+\nu_{-}\mu_{-})/(\nu_{+}+\nu_{-})$.~\cite{valleau_jcp_80}
The underlying model is the ``primitive'' model of electrolytes where the ions are modeled as charged hard spheres (diameters $d_{+}=d_{-}=d=0.3$ nm in this study) immersed in an implicit solvent of dielectric constant $\epsilon$. 
The infinite tube is modeled via periodic boundary conditions applied in the $z$ direction.
The wall of the nanopore is hard, while the interaction of the ions with the uniform surface charge is computed from a fit to interaction with partial point charges on a high-resolution grid.  

Both methods provide the electrical potential profiles, $\Phi(r)$, the concentration profiles, $c_{i}(r)$, and the average numbers of ions in a unit length of the pore, $N_{i}$.

\subsection{Extension to multivalent electrolytes}

\begin{figure}[t!]
\begin{center}
\includegraphics[width=0.46\textwidth]{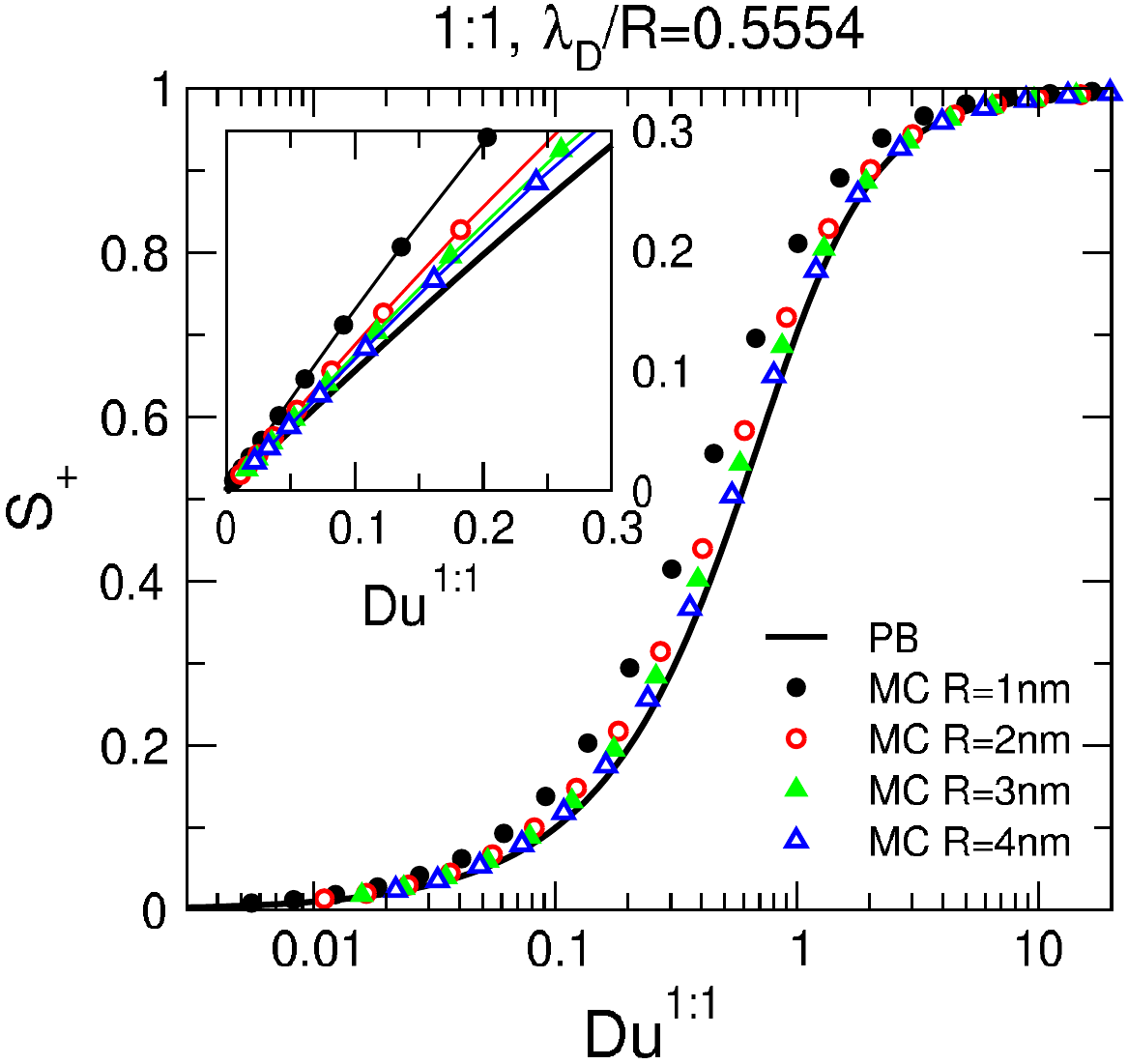}
\end{center}
\caption{Selectivity curves obtained by scanning the surface charge ($-3\leq \sigma \leq - 0.001$ $e$/nm$^{2}$) as functions of the Dukhin number for a 1:1 electrolyte ($\mathrm{Du}=\mathrm{Du}^{1:1}$) for a fixed $\lambda_{\mathrm{D}}/R=0.5554$. 
Here and in following figures, the curves have been obtained by numerically solving the nonlinear PB equation (Eq.~\ref{eq:pb}), while the symbols show results of MC simulations for hard-sphere ions. 
Here, MC simulations have been performed for concentrations that give  $\lambda_{\mathrm{D}}/R=0.5554$ for pore radii $R=1$, $2$, $3$, and $4$ nm  (these concentrations are $c=0.3$, $0.075$, $0.035$, and $0.019$ M). 
Here, and in following figures, the insets show the results on a linear scale of the $\mathrm{Du}^{1:1}$ axis for small values of $\mathrm{Du}^{1:1}$.
The thin lines connecting the symbols are to guide the eyes regarding the slopes.
}
\label{fig1}
\end{figure}

In this work, we generalize the Dukhin number to electrolytes containing multivalent ions on the basis of the PB equation.
For this, we plot $S_{+}$ against $\mathrm{Du}^{1:1}$ on the linear scale of $\mathrm{Du}^{1:1}$ for small values of $\mathrm{Du}^{1:1}$ (see the insets of Figs.~\ref{fig1} and \ref{fig2}).
The insets imply that $S_{+}=\mathrm{Du}^{1:1}$ for small values of $\mathrm{Du}^{1:1}$.
This is the region where $\sigma$ is small, the pore is poorly selective, and $\Phi(r)$ is close to zero.
In this regime, the Boltzmann-factor in Eq.~\ref{eq:c_i} can be linearized providing

\newpage
\begin{widetext}
\begin{equation}
\begin{aligned}
 S_{+} &= \dfrac{ D_{+}z_{+}c_{+}^{\mathrm{b}}\left[ \bigintss\limits_{0}^{R} \left( 1-\dfrac{z_{+}e\Phi(r)}{kT} \right) 2\pi r \mathrm{d}r \right] - 
                  D_{-}|z_{-}|c_{-}^{\mathrm{b}}\left[ \bigintss\limits_{0}^{R} \left( 1+\dfrac{|z_{-}|e\Phi(r)}{kT} \right) 2\pi r \mathrm{d}r \right] }
                { D_{+}z_{+}c_{+}^{\mathrm{b}}\left[ \bigintss\limits_{0}^{R} \left( 1-\dfrac{z_{+}e\Phi(r)}{kT} \right) 2\pi r \mathrm{d}r \right] + 
                  D_{-}|z_{-}|c_{-}^{\mathrm{b}}\left[ \bigintss\limits_{0}^{R} \left( 1+\dfrac{|z_{-}|e\Phi(r)}{kT} \right) 2\pi r \mathrm{d}r \right] }  = \\
       &= \dfrac{ D_{+}z_{+} c_{+}^{\mathrm{b}}\left[ A - \dfrac{z_{+}e\alpha}{kT}  \right] -   
                  D_{-}|z_{-}|c_{-}^{\mathrm{b}}\left[   A +\dfrac{|z_{-}|e\alpha}{kT}  \right] }
                { D_{+}z_{+}c_{+}^{\mathrm{b}}\left[    A-\dfrac{z_{+}e\alpha}{kT} \right] + 
                  D_{-}|z_{-}|c_{-}^{\mathrm{b}}\left[   A+\dfrac{|z_{-}|e\alpha}{kT} \right] }      
 \label{eq:sderivationlong}               
\end{aligned} 
\end{equation}
\end{widetext}
where $\alpha = \int_{0}^{R} \Phi(r) 2\pi r\mathrm{d}r$ and $A=R^{2}\pi$ were introduced and $-|z_{-}|$ were used instead of $z_{-}$.
From electroneutrality in the bulk it follows that $z_{+}c_{+}^{\mathrm{b}}=|z_{-}|c_{-}^{\mathrm{b}}$.  
Eq.~\ref{eq:sderivationlong}, therefore, can be expressed as 
\begin{equation}
 S_{+} = \dfrac{\left(D_{+}-D_{-}\right)-\left( D_{+}z_{+} + D_{-}|z_{-}| \right) \dfrac{e\alpha}{AkT} } 
               {\left(D_{+}+D_{-}\right) - \left( D_{+}z_{+} - D_{-}|z_{-}| \right) \dfrac{e\alpha}{AkT} }.
\label{eq:sel3}
 \end{equation} 
We can obtain a closed formula for $\alpha$ if we integrate the linearized form of Eq.~\ref{eq:pb},
\begin{equation}
 \frac{1}{r} \frac{\mathrm{d}}{\mathrm{d} r} \left( r \frac{\mathrm{d} \Phi(r)}{\mathrm{d} r} \right) = \frac{1}{\lambda_{\mathrm{D}}^{2}} \Phi(r) 
 \label{eq:pblin}
\end{equation} 
yielding
\begin{equation}
\bigintsss\limits_{0}^{R} \frac{1}{r} \frac{\mathrm{d}}{\mathrm{d} r} \left( r \frac{\mathrm{d} \Phi(r)}{\mathrm{d} r} \right) 2\pi r\mathrm{d}r = \frac{1}{\lambda_{\mathrm{D}}^{2}} \int\limits_{0}^{R} \Phi(r) 2\pi r \mathrm{d}r .
 \label{eq:pb2}
\end{equation}
The right hand side is $\alpha / \lambda_{\mathrm{D}}^{2}$, while the left hand side is 
\begin{equation}
2\pi \left[r \frac{\mathrm{d}\Phi(r)}{\mathrm{d}r} \right]_{0}^{R} = 2\pi R \left(  \frac{\mathrm{d}\Phi(r)}{\mathrm{d}r} \right)_{r=R} = \frac{2\pi R}{\epsilon_{0}\epsilon } \sigma 
\end{equation} 
so 
\begin{equation}
 \alpha = \frac{\sigma 2\pi \lambda_{\mathrm{D}}^{2} R }{\epsilon_{0}\epsilon}.
\end{equation} 
Comparing this result to Eq.~\ref{eq:du1}, we obtain the relation 
\begin{equation}
 \frac{e\alpha}{AkT}= - \mathrm{Du}^{0}.
\end{equation} 
Substituting this into Eq.~\ref{eq:sel3}, we obtain that
\begin{equation}
 S_{+} =  \dfrac{\left(D_{+}-D_{-}\right)+\left( D_{+}z_{+} + D_{-}|z_{-}| \right) \mathrm{Du}^{0}} { \left(D_{+}+D_{-}\right) + \left( D_{+}z_{+} - D_{-}|z_{-}| \right) \mathrm{Du}^{0} } ,
 \label{eq:s_vs_du}
\end{equation} 
where $\mathrm{Du}^{0}$ is given by Eq.~\ref{eq:du1}.

If we introduce the notation $D^{*}=D_{+}/D_{-}$, Eq.~\ref{eq:s_vs_du} simplifies to 
\begin{equation}
 S_{+} =  \dfrac{\left(D^{*}-1\right) + \left(D^{*}z_{+} + |z_{-}| \right) \mathrm{Du}^{0}} 
                {\left(D^{*} + 1\right) + \left( D^{*}z_{+} -  |z_{-}| \right) \mathrm{Du}^{0} } .
 \label{eq:genDu_withD}
\end{equation} 
This derivation is valid in the linearized PB regime, which is the regime of small $\mathrm{Du}^{0}$.
Therefore, if we introduce the generalized Dukhin number as 
\begin{equation}
\mathrm{Du} =  \dfrac{\left(D^{*}-1\right) + \left( D^{*}z_{+} + |z_{-}| \right) \mathrm{Du}^{0}} 
                     {\left(D^{*}+1\right) + \left( D^{*}z_{+} -  |z_{-}| \right) \mathrm{Du}^{0} } ,
 \label{eq:genDu}
\end{equation} 
we can write that
\begin{equation}
 S_{+}=\lim_{\mathrm{Du}^{0}\to 0} \mathrm{Du}\left(\mathrm{Du}^{0}\right).
\end{equation} 
Although this analytical relationship is valid only in the $\mathrm{Du}^{0}\to 0$ limit, it is reasonable to assume that a single $f\left[\mathrm{Du}\left(\mathrm{Du}^{0}\right)\right]$ function (a ``master curve'') exists for the whole range of $\mathrm{Du}^{0}$ that approximates the actual numerical solution satisfactorily.

\subsection{Simplifications}
\label{subsec:simpl}

Our equilibrium calculations provide the average numbers of ions in the pore determined by the spatial distributions.
They do not give an account of the mobilities of the ions that are characterized by the $D^{*}$ parameter in the resulting equations (Eqs.~\ref{eq:genDu_withD} and \ref{eq:genDu}).

So, we can make our discussion independent of diffusion constants if we assume that $D_{+}=D_{-}$, which means that $D^{*}=1$.
In this case, Eq.~\ref{eq:genDu} simplifies to
\begin{equation}
 \mathrm{Du} =  \dfrac{\left( z_{+} + |z_{-}| \right) \mathrm{Du}^{0}} { 2 + \left( z_{+} - |z_{-}| \right) \mathrm{Du}^{0} } .
 \label{eq:genDu_simp}
\end{equation}  
We will discuss this scaling parameter from now on and we call it a generalized Dukhin number, or just simply a Dukhin number.
We emphasize that $\mathrm{Du}^{0}$ should be computed from Eq.~\ref{eq:du1}, not from Eq.~\ref{eq:dukhin}, because $\lambda_{\mathrm{D}}$ carries a dependence on the ionic valences.

\begin{figure}[t!]
\begin{center}
\includegraphics[width=0.46\textwidth]{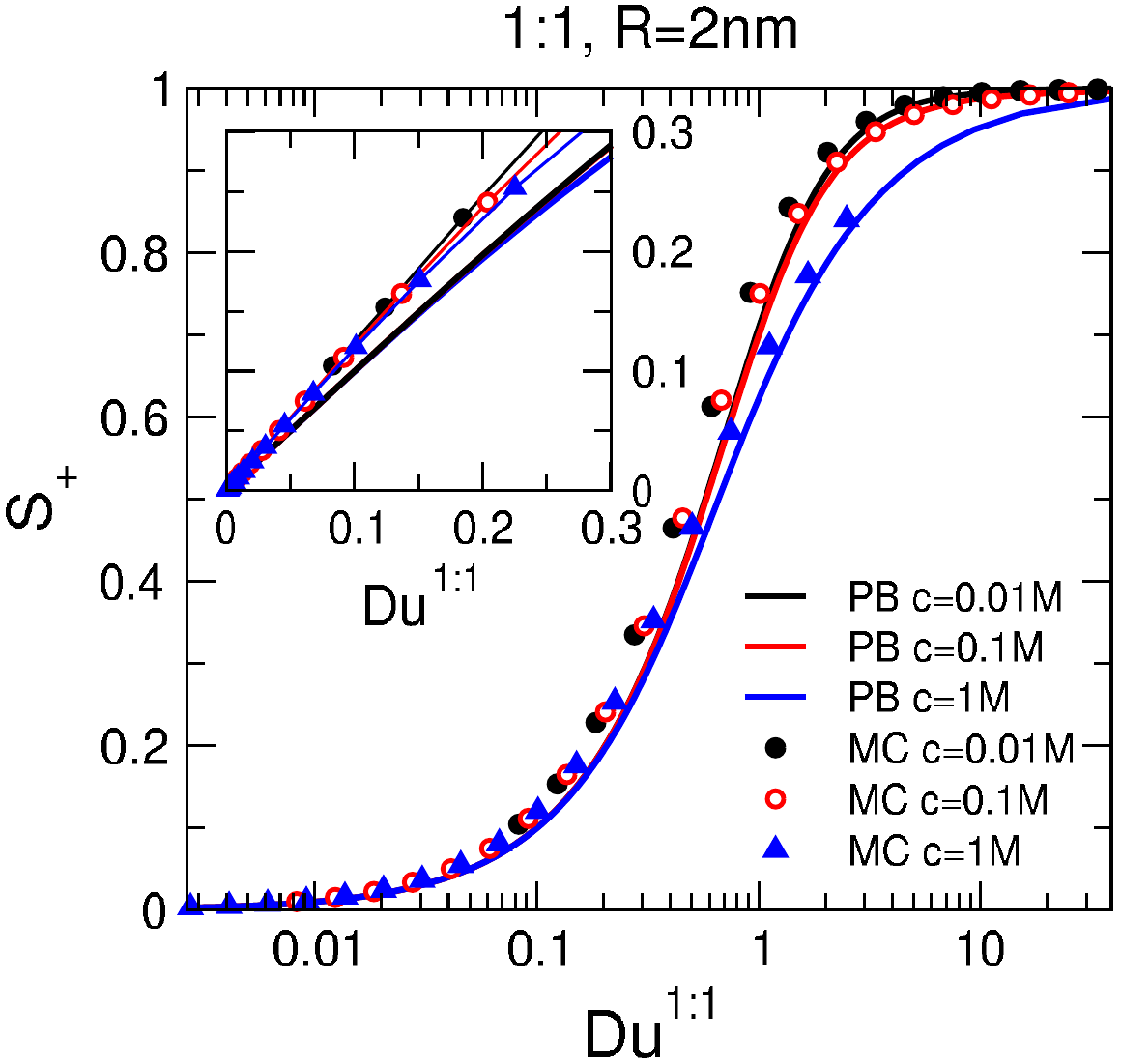}
\end{center}
\caption{Selectivity curves obtained by scanning the surface charge ($-3\leq \sigma \leq - 0.001$ $e$/nm$^{2}$) as functions of the Dukhin number for 1:1 electrolytes ($\mathrm{Du}=\mathrm{Du}^{1:1}$) for a fixed $R=2$ nm at different concentrations. 
Lines, symbols, and inset have the same meaning as in Fig.~\ref{fig1}.
}
\label{fig2}
\end{figure}

Note that we could have derived this expression by defining selectivity as
\begin{equation}
 S_{+}=\frac{|J_{+}|/D_{+}-|J_{-}|/D_{-}}{|J_{+}|/D_{+}+|J_-|/D_{-}} .
 \label{eq:selnew}
\end{equation} 
This way, only the concentration-dependent part of the NP equation contributes to selectivity.

Since the Dukhin number was originally proposed with 1:1 electrolytes in mind~\cite{bocquet_chemsocrev_2010}, it is instructive to discuss the 1:1 case in detail.
For the 1:1 electrolytes ($z_{+}=1$ and $z_{-}=-1$), Eq.~\ref{eq:genDu_simp} simplifies to $S_{+}=\mathrm{Du}^{0}=\mathrm{Du}^{1:1}$.
Fig.~\ref{fig1} shows that the PB solution provides the same curve for different combinations of $c$ and $R$ that produce the same $\lambda_{\mathrm{D}}/R$ ratio. 
Fig.~\ref{fig2} shows that an $f\left[\mathrm{Du}^{1:1}\right]$ master curve can also be obtained as soon as the concentration is not too large ($\lambda_{\mathrm{D}}/R$ is not too small).

The insets of these figures show that selectivity is a linear function of $\mathrm{Du}^{1:1}$ for small values of $\mathrm{Du}^{1:1}$.
If the slope of the $S_{+}$ vs.\ $\mathrm{Du}^{1:1}$ curve in the origin is the same for different conditions, there is a chance that $S_{+}$ will depend on $\mathrm{Du}^{1:1}$ via the same function for different conditions.
If this assumption is valid, $\mathrm{Du}^{1:1}$ is an appropriate scaling parameter.
The logic of the derivation is that the scaling parameter follows from a limiting case, while the validity of the scaling for the whole parameter range is justified by actual numerical calculations.

\begin{figure}[t!]
\begin{center}
\includegraphics[width=0.46\textwidth]{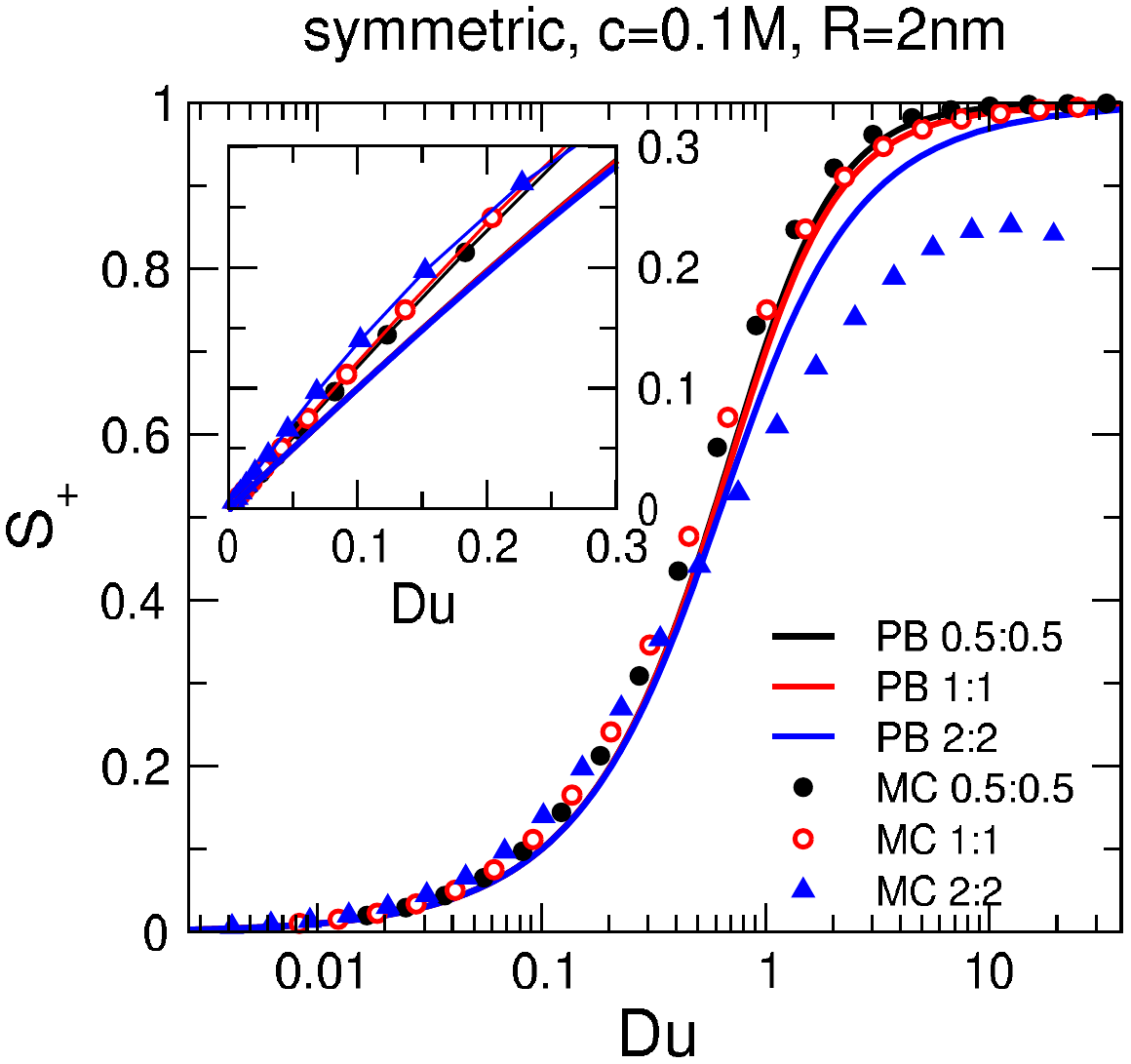}
\end{center}
\caption{Selectivity curves obtained by scanning the surface charge ($0.001 \leq \sigma \leq 3$ $e$/nm$^{2}$) as functions of the generalized Dukhin number (Eq.~\ref{eq:genDu}) for  symmetric electrolytes ($z_{+}=|z_{-}|$) for $R=2$ nm and $c=0.1$ M. 
The $z=0.5$ case represents a model for a weak ionic coupling.
Lines, symbols, and inset have the same meaning as in Fig.~\ref{fig1}.
}
\label{fig3}
\end{figure}

For symmetric electrolytes, Eq.~\ref{eq:genDu_simp} simplifies to
\begin{equation}
 \mathrm{Du}^{z:z} =  z \mathrm{Du}^{0},
 \label{eq:duzz}
\end{equation} 
where $z=z_{+}=|z_{-}|$.
The inset of Fig.~\ref{fig3} shows that the scaling also works for symmetric electrolytes different from 1:1 in the $\mathrm{Du}\to 0$ limit.
The main panel of Fig.~\ref{fig3} shows that the scaling also works reasonably in the whole range of $\mathrm{Du}$.

\begin{figure*}[t!]
\begin{center}
\includegraphics[width=0.7\textwidth]{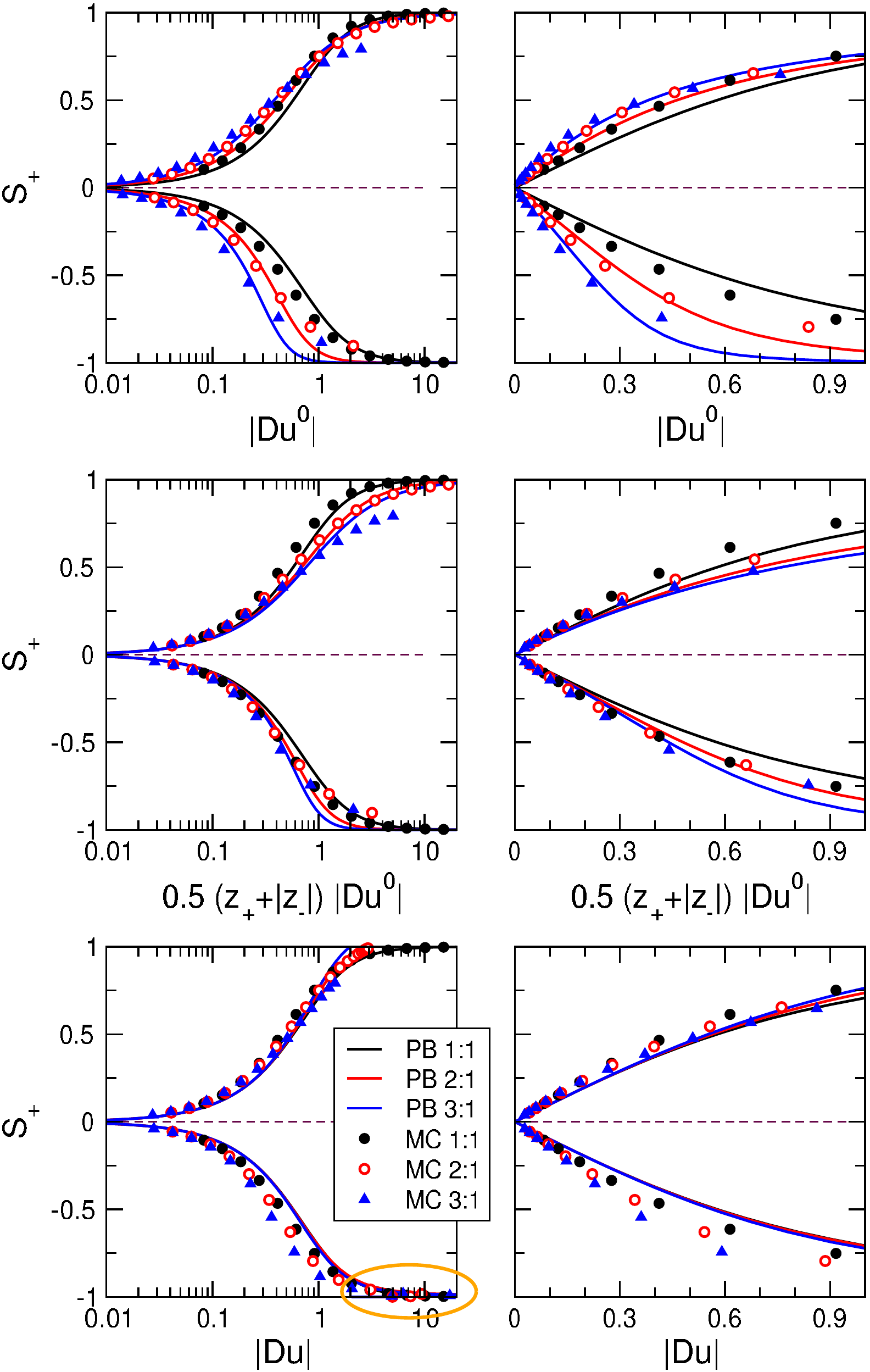}
\end{center}
\caption{Selectivity curves obtained by scanning the surface charge ($0.001 \leq |\sigma| \leq 3$ $e$/nm$^{2}$) as functions of $|\mathrm{Du}^{0}|$ (top row), $\sfrac{1}{2}\; (z_{+}+|z_{-}|)\; |\mathrm{Du}^{0}|$ (middle row), and $|\mathrm{Du}|$ (bottom row) for $R=2$ nm and $c=0.01$ M.
Left and right columns show the same results on a logarithmic and a linear scale, respectively.
Lines and symbols have the same meaning as in Fig.~\ref{fig1}.
The orange oval in the bottom-left panel indicates the region where divergence of the $\mathrm{Du}\left(\mathrm{Du}^{0}\right)$ function occurs.
This divergence behavior is discussed in Fig.~\ref{fig6} in detail.
}
\label{fig4}
\end{figure*}

For asymmetric electrolytes, the bottom-right panel of Fig.~\ref{fig4} shows that the scaling works also for asymmetric electrolytes in the $\mathrm{Du}\to 0$ limit.
The bottom-left panel shows that the scaling also works reasonably in the whole range of $\mathrm{Du}$.
Because $\mathrm{Du}$ is a rational function of $\mathrm{Du}^{0}$, $\mathrm{Du}$ diverges if the denominator approaches zero, namely, for $\mathrm{Du}^{0}\to -2/(z_{+}-|z_{-}|)$.
This is not really an elegant behavior for a scaling parameter. 
Problems caused by this behavior and possible solutions will be discussed in the next section.

\section{Results}

Figures~\ref{fig1}--\ref{fig3} and the bottom row of Fig.~\ref{fig4} show that the scaling works not only in the $\mathrm{Du}\to 0$ limit, but also for the whole $\mathrm{Du}$ range, especially, for intermediate values of $\mathrm{Du}$.
The sigmoid curves collapse onto a single curve with a reasonable accuracy. 
Deviations occur at large values of $\mathrm{Du}$ corresponding to large selectivities if we change $\lambda_{\mathrm{D}}/R$ or ionic valences.

\subsection{1:1 electrolytes}
In the framework of the PB theory, we obtain the same curve for a fixed value of $\lambda_{\mathrm{D}}/R$ with different combinations of $R$ and $c$ for 1:1 electrolytes (Fig.~\ref{fig1}).
This collapse at a fixed $\lambda_{\mathrm{D}}/R$ is also valid for other electrolytes.

Fig.~\ref{fig1} also shows the $R$ dependence of the MC results while the $\lambda_{\mathrm{D}}/R$ ratio is kept constant.
The MC points slightly differ from the PB curve due to the finite size of the ions and electrostatic correlations beyond the mean-field approximation. 
As the pore radius is increased, the MC points get closer to the PB curve because $d_{i}$ gets smaller relative to $R$ and the importance of the hard-sphere exclusion effects decreases.
This is seen both on the whole sigmoid curve and the linear regime in the inset.
The differences due to electrostatic correlations, however, are present even at large $R$.

Fig.~\ref{fig2} shows that the deviation from the ``master curve'' appears at larger concentrations if $R$ is fixed using both methods.
Because the PB theory captures the $c$ dependence given by the MC simulations pretty well, we can conclude that the behavior of the 1:1 electrolyte is chiefly governed by mean-field effects.

\subsection{Symmetric $z:z$ electrolytes}
Fig.~\ref{fig3} shows that deviations also appear by changing $z=z_{+}=|z_{-}|$ of a symmetric electrolyte for fixed $c$ and $R$.
Deviations increase with increasing values of $z$, namely, when electrostatic correlations get stronger.
The PB and MC results agree quite well for $z=0.5$ and $1$.

\begin{figure}[t!]
\begin{center}
\includegraphics[width=0.46\textwidth]{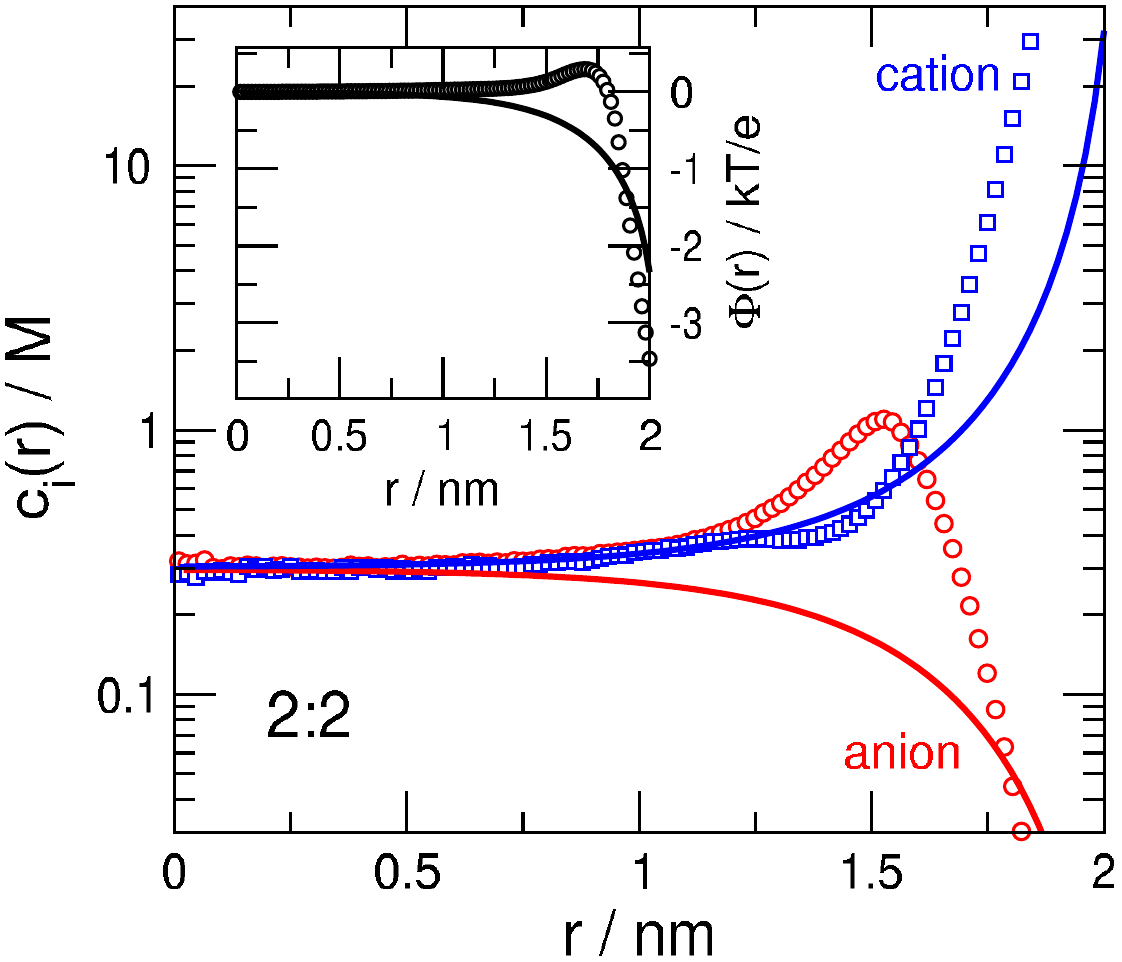}
\end{center}
\caption{Radial concentration profiles for a 2:2 electrolyte for $R=2$ nm,  $c=0.3$ M, and $\sigma =-2$ $e$/nm$^{2}$. Symbols and lines show MC and PB results, respectively. The insets show potential profiles in $kT/e$ unit. 
}
\label{fig5}
\end{figure}

For the 2:2 system, however, a considerable deviation is observed between the PB and MC data.
As $\mathrm{Du}$ is increased to larger values, the MC points are shifted towards larger values of $\mathrm{Du}$ and the sigmoid does not converge to $1$ as $\mathrm{Du}\to\infty$.
This is the result of charge inversion when electrostatic correlations (beyond the mean-field level) are so strong that cations overcharge the wall~\cite{he_jacs_2009}  and a layer of excess anions occurs in the pore (Fig.~\ref{fig5}). 

This results in an increased concentration of the anions in the pore because the cations correlate strongly with the anions and carry them in the pore with themselves. 
This leads to an ``anion leakage'' that prevents cation selectivity from being perfect ($S_{+}=1$) even if $\sigma$ is very large.

Charge inversion is absent in the PB theory: the concentration and potential profiles are monotonic (see lines in Fig.~\ref{fig5}) as opposed to the MC curves where the anion layer causes non-monotonic behavior (see symbols in Fig.~\ref{fig5}).
The PB curves for different $z:z$ electrolytes collapse onto a single curve only if we use the rescaled Dukhin number, $\mathrm{Du}^{z:z}=z\mathrm{Du}^{0}$ (Eq.~\ref{eq:duzz}) instead of $\mathrm{Du}^{1:1}$.

\begin{figure}[t!]
\begin{center}
\includegraphics[width=0.46\textwidth]{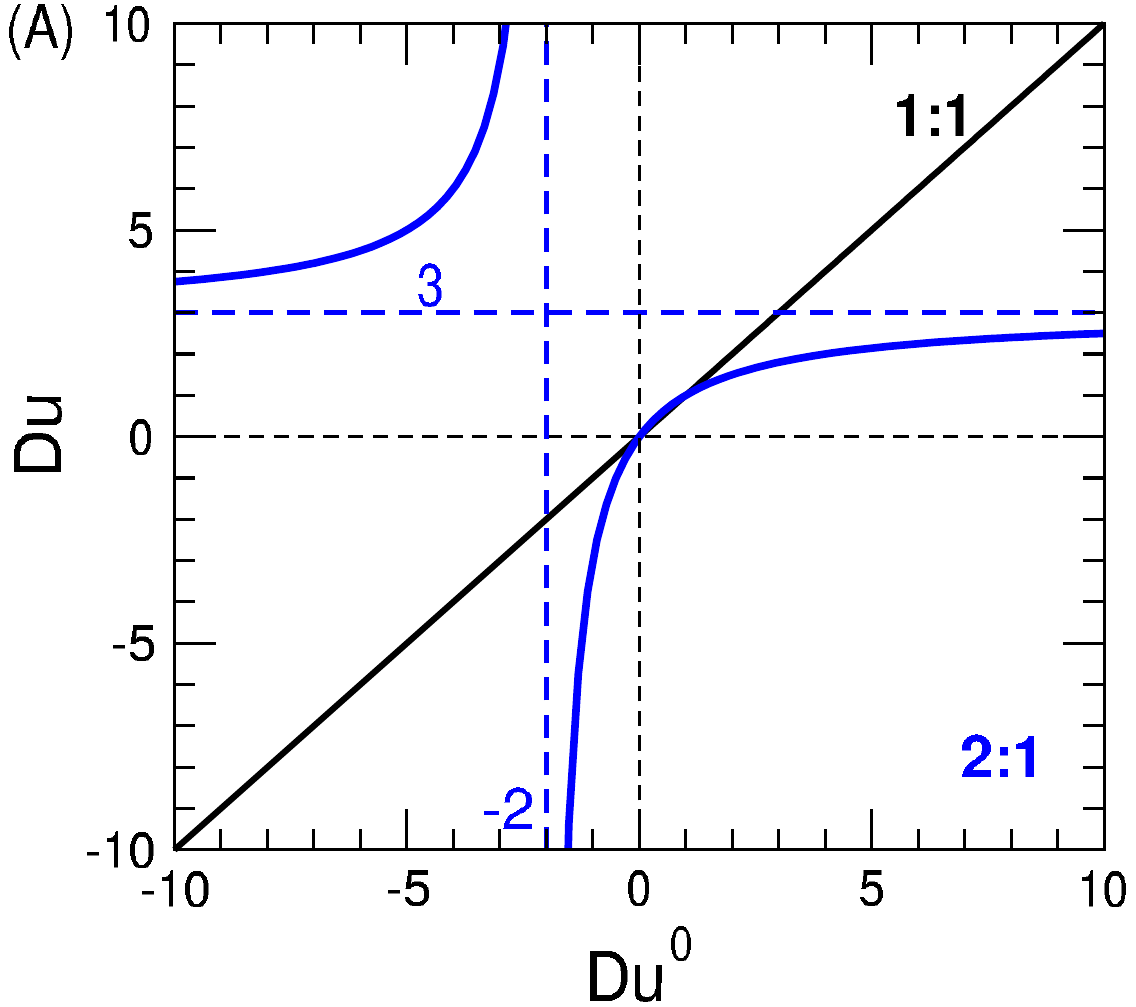}\vspace{0.3cm}
\includegraphics[width=0.46\textwidth]{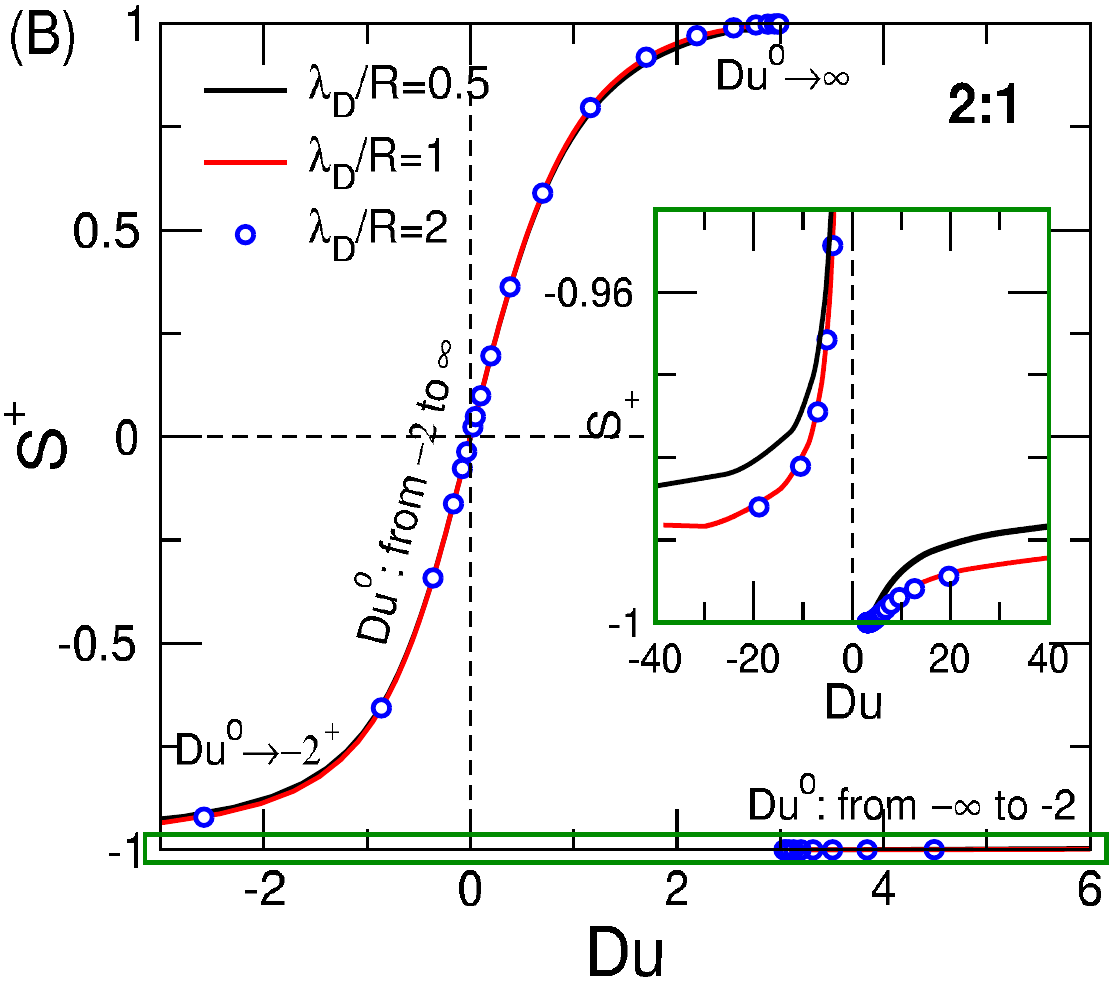}
\end{center}
\caption{(A) The $\mathrm{Du}$ parameter as a function of the $\mathrm{Du}^{0}$ parameter for a 2:1 electrolyte. (B) Selectivity curves as functions of $\mathrm{Du}$ for different values of $\lambda_{\mathrm{D}}/R$ as obtained from the PB theory for a 2:1 system.
The inset zooms at the region where the divergence of $\mathrm{Du}$ occurs (large $\sigma>0$ and large anion selectivity $S_{+}\sim -1$).
The region framed with a green rectangle in the main panel is shown in the inset.
}
\label{fig6}
\end{figure} 

\subsection{Asymmetric electrolytes}
For asymmetric electrolytes, the $\mathrm{Du}^{0}\to \mathrm{Du}$ transformation is less simple: $\mathrm{Du}$ is obtained from $\mathrm{Du}^{0}$ via a rational function (Eq.~\ref{eq:genDu_simp}) with the valences $z_{+}$ and $|z_{-}|$ in the coefficients.
In this case, other options for the scaling parameter need to be discussed.

Fig.~\ref{fig4} shows the selectivity curves as functions of various parameters.
The asymmetric electrolytes behave differently at the negative and positive signs of the surface charge density.
In this case, therefore, we need to plot the $S_{+}$ vs.\ $\mathrm{Du}$ function for the whole range $-\infty <\mathrm{Du}<\infty$.
In order to be able to use a logarithmic scale, we plot the results as functions of the absolute values of the respective parameters.

The top row shows the results as functions of $\mathrm{Du}^{0}$.  
Although $\mathrm{Du}^{0}$ carries an explicit $\sigma$, $c$, $R$, and $z_{i}$ dependence, it does not produce a scaling in between different $z_{+}:z_{-}$ electrolytes.
If we remain in the framework of a given electrolyte (fixed $z_{+}$ and $z_{-}$), however, we can use $\mathrm{Du}^{0}$ as a scaling parameter at the price that we lose the $\lim_{\mathrm{Du}^{0}\to 0} S_{+} = \mathrm{Du}^{0}$ limiting case.

It is tempting to use the Taylor series expansion of $\mathrm{Du}$ as a function of $\mathrm{Du}^{0}$.
Our calculations (results not shown), however, did not really give reasonable results even up to quite high-order terms.
As a consequence, we deemed this to be impractical, although it is worth keeping in mind as an option.

The first-order term of the series expension, however, gives an approximate solution, 
\begin{equation}
 \mathrm{Du}\approx \frac{z_{+}+|z_{-}|}{2} \mathrm{Du}^{0} ,
\end{equation} 
that can also be considered as a limiting case as $\mathrm{Du}^{0}\to 0$.
The middle row of Fig.~\ref{fig4} shows the curves as functions of this parameter.
It is apparent that the $\sfrac{1}{2}\, (z_{+}+|z_{-}|)$ multiplier fixes the slopes at the origin, but the deviations between the various $z_{+}:z_{-}$ cases are larger compared to the bottom row, which shows the curves as functions of $\mathrm{Du}$. 

The bottom row of Fig.~\ref{fig4} shows that the scaling works nicely using $\mathrm{Du}$ as the scaling parameter.
As it was already indicated in section \ref{subsec:simpl}, $\mathrm{Du}$ as defined in Eq.~\ref{eq:genDu_simp} diverges when the denominator approaches zero, namely, as $\mathrm{Du}^{0}\to -2/(z_{+}-|z_{-}|)$.
The divergence is absent in the case of the $\mathrm{Du}^{0}$ and $\sfrac{1}{2}\, (z_{+}+|z_{-}|)\, \mathrm{Du}^{0}$ parameters at the price of weaker accuracy.
The region of divergence is shown by an orange ellipse in Fig.~\ref{fig4} and investigated in Fig.~\ref{fig6} in detail.

Using a 2:1 electrolyte as an example, we show the $\mathrm{Du}\left(\mathrm{Du}^{0}\right)$ function in Fig.~\ref{fig6}a.
Its behavior can be described by the limits
\begin{equation}
 \lim_{\mathrm{Du}^{0}\to-2^{+}} \mathrm{Du}\left(\mathrm{Du}^{0}\right)=-\infty ,
\end{equation}
\begin{equation}
 \lim_{\mathrm{Du}^{0}\to-2^{-}} \mathrm{Du}\left(\mathrm{Du}^{0}\right)=\infty  ,
\end{equation}
\begin{equation}
 \lim_{\mathrm{Du}^{0}\to \pm \infty} \mathrm{Du}\left(\mathrm{Du}^{0}\right)= 3 .
\end{equation}
Figure \ref{fig6}b shows that the cation selective branch ($S_{+}>0$) belongs to $\mathrm{Du}\in (0,3)$ corresponding to $\mathrm{Du}^{0}\in (0,\infty)$. The anion selective branch, however, is split into two parts: (1) for $\mathrm{Du}\in (-\infty ,0)$ corresponding to $\mathrm{Du}^{0}\in (-2 , 0)$ selectivity changes in the range $S_{+}\in (-0.99 , 0)$, while (2)  for $\mathrm{Du}\in (3, \infty)$ corresponding to $\mathrm{Du}^{0}\in (-\infty, -2)$ selectivity changes in the range $S_{+}\in (-1, -0.99)$.

This divergence problem appears at different conditions depending on the parameters $z_{+}$, $z_{-}$, and $D^{*}$. 
In the 3:1 system, for example, the divergence occurs as $\mathrm{Du}^{0}\to -1$.
In the case of different diffusion constants (Eq.~\ref{eq:genDu}), $\mathrm{Du}$ diverges as $\mathrm{Du}^{0}\to -(D^{*}+1)/(D^{*}z_{+} -  |z_{-}|)$ even in 1:1 systems.
Despite of the noncontinuous behavior of the $S_{+}$ vs.\ $\mathrm{Du}$ function, scaling works: the PB curves for larger values of $\lambda_{\mathrm{D}}/R$ collapse onto each other (see red curves and blue symbols in Fig.~\ref{fig6}b). 

As far as the PB vs.\ MC comparison is concerned, charge inversion appears in 2:1 and 3:1 electrolytes as well.
In the case of the 3:1 system, for example, the ionic correlations are so strong that an anomalous behavior occurs at large $\sigma$ values.
This behavior, however, is not a turnover as in the case of the 2:2 system, but a phase transition.
The ``primitive'' model of electrolytes used in this study has a vapor-liquid (or dilute-dense) phase transition if the ionic coupling (the Coulomb interaction between cations and anions in contact relative to $kT$) is large enough.~\cite{orkoulas_jcp_1994,cheong_jcp_2003}
This phase transition from a dilute to a dense phase appears at a lower chemical potential than in the bulk, so we can conclude that this phase separation is induced by pore confinement and/or surface charge~\cite{pizio_jcp_2004}.
 
\section{Summary}

The scaling behavior presented in this study relates a device function to a scaling parameter, where both of these quantities are rational functions of more basic parameters ($J_{i}$, $z_{i}$, $c$, $R$, $H$, $\sigma$, $U$).
The device function is constructed from output quantities that are obtained as results of a measurement or calculation ($J_{i}$).
The scaling parameter is constructed from quantities that can be tuned relatively easily during nanopore fabrication ($R$, $H$, $\sigma$) or the experimental setup ($z_{i}$, $c$, $U$).

While this work considers the $H\to \infty$ limit, in our previous work~\cite{sarkadi_jcp_2021} we showed for 1:1 electrolytes that the scaling parameter for finite pores approaching the nanohole ($H\to 0$) limit is a modified Dukhin number defined as 
\begin{equation}
 \mathrm{mDu}^{1:1} = \mathrm{Du}^{1:1} \frac{H}{\lambda_{\mathrm{D}}}.
\end{equation} 
We also showed that $\mathrm{mDu}^{1:1}$ is an appropriate scaling parameter for rectification in bipolar nanopores.~\cite{fertig_ms_2022} 
Generalization of $\mathrm{mDu}$ for multivalent electrolytes will be reported soon.

An important difference between the finite nanopores and the infinite one is that finite pores do not have to be charge neutral because charge accumulation near the membrane at the two sides near the pore entrances may screen the net charge inside the pore.
This leads to a polarization of charge densities in the axial dimension as well thus leading to scaling phenomena dependent on the pore length.
Breakdown of charge neutrality, however, may appear even in long pores if the pore is surrounded by a dielectric material whose polarization charge contributes to maintaining effective charge neutrality.~\cite{levy_jcis_2020,desouza_pre_2021,green_jcp_2021b} 
Dielectric boundaries are absent in this work and charge neutrality is imposed in our calculations. 
In the PB theory, Eq.~\ref{eq:sigma} ensures that the electric field is zero outside the pore, while in the MC simulations, charge neutrality is satisfied in every simulation step because we insert/delete neutral ion clusters.
Although breakdown of charge neutrality is an important effect, we ignore entrance effects in this work that helps us in staying focused.

The sigmoid curve includes two important limiting cases: (1) micropores where selectivity is small and bulk conduction dominates, and (2) highly charged nanopores where coions are excluded, selectivity is large and surface conduction dominates.
These cases allow approximations in theoretical considerations and lead to elegant analytical formulas.~\cite{green_jcp_2021}

Scaling is a phenomenon resulting from an analytical solution. The numerical mean-field solution of the PB theory provides more accurate scaling than MC simulations because it contains approximations that are not present in MC simulations that provide exact solution for the model. Deviations between the PB and MC results for scaling, therefore, indicate the importance of ionic correlations beyond the mean-field level.

\section*{Acknowledgements}
\label{sec:ack}

We gratefully acknowledge  the financial support of the National Research, Development, and Innovation Office -- NKFIH K137720 and the TKP2021-NKTA-21.
We are grateful to Dirk Gillespie, Tam\'as Krist\'of, and Zolt\'an Hat\'o for inspiring discussions.

\end{document}